\documentclass[aps,prl,article,twocolumn,preprintnumbers,amsmath,amssymb,superscriptaddress]{revtex4-2}

\usepackage{graphicx}  
\usepackage{dcolumn}   

\usepackage{bm}        
\usepackage{amssymb}   
\usepackage{amsmath}
\usepackage{mathrsfs}
\usepackage{epigraph}
\usepackage{braket}
\usepackage{gensymb}
\usepackage{lmodern}
\usepackage{ tipa }
\usepackage{bbold}
\usepackage{esint}
\usepackage{mathdots}
\usepackage{xcolor}
\usepackage{appendix}
\usepackage{natbib}
\usepackage{multirow}
\usepackage{float}

\usepackage{comment}
\usepackage[colorlinks=true, linkcolor=blue, urlcolor=blue, citecolor=blue]{hyperref}

\newcommand{\rvss}[1]{{{#1}}}
\begin{document}
\title{Light Control of Triplet Pairing in Correlated Electrons with Mixed-Sign Interactions} 	
\author{Zecheng Shen}
\affiliation{Department of Chemistry, Emory University, Atlanta, GA 30322, USA}
\author{Chendi Xie}
\affiliation{Department of Physics and Astronomy, Clemson University, Clemson, SC 29634, USA}
\affiliation{Department of Chemistry, Emory University, Atlanta, GA 30322, USA}
\author{Wei-Chih Chen}
\affiliation{Department of Physics and Astronomy, Clemson University, Clemson, SC 29634, USA}
\author{Yao Wang}

\email[Correspondence should be addressed to \href{mailto:yao.wang@emory.edu}{yao.wang@emory.edu}]{}
\affiliation{Department of Chemistry, Emory University, Atlanta, GA 30322, USA}
\date{\today}
\begin{abstract} \textbf{Abstract:} 
    Spin-triplet superconductivity is a key platform for topological quantum computing, yet its experimental realization and control in solid-state materials remain a significant challenge. For this purpose, we propose an ultrafast optical strategy to manipulate spin-triplet superconductivity by leveraging $p$-wave pairing instabilities in the extended Hubbard model, a framework applicable to transition-metal oxides. Utilizing Floquet engineering, we demonstrate that transient flipping of the effective spin-exchange interaction can enhance $p$-wave pairing correlations under linearly polarized optical pulses. Furthermore, we reveal that this emergent spin-triplet pairing in strongly correlated systems can be selectively switched by an orthogonal optical pulse. This work provides a pathway for stabilizing and controlling spin-triplet superconductivity in correlated materials. 
\end{abstract}

\maketitle

\section{Introduction}
Spin-triplet superconductivity is widely recognized as a promising avenue for topological quantum computing\,\cite{sarma2015majorana,nayak2008non}, offering a robust platform for hosting non-Abelian excitations\,\cite{read2000paired,ivanov2001non,kitaev2001unpaired}. However, despite its theoretical appeal, experimental realization remains an ongoing challenge. Sr$_2$RuO$_4$ was long considered the leading candidate, particularly for a chiral $p+ip$ pairing state\,\cite{maeno1994superconductivity,rice1995sr2cuo4,baskaran1996why,mackenzie2003the, maeno2011evaluation, nelson2004odd}, but recent experimental findings have cast significant doubt on its pairing symmetry\,\cite{pustogow2019constraints,ishida2020reduction,chronister2021evidence}. Other materials, such as UTe$_2$\,\cite{ran2019nearly,ran2020enhancement}, UPt$_3$\,\cite{tou1996odd}, and K$_2$Cr$_3$As$_3$\,\cite{yang2021spin} have been proposed as potential candidates for spin-triplet superconductors, yet unambiguous experimental validation of spin-triplet pairing remain elusive.

Beyond specific materials, theoretical and numerical studies have revealed that spin-triplet pairing can naturally emerge in strongly correlated models with mixed-sign interactions. In particular, the one-dimensional (1D) extended Hubbard model (EHM), where a repulsive on-site interaction $U$ coexists with an attractive nearest-neighbor interaction $V$, has been shown to support nodal pairing symmetry. Under appropriate doping and parameter regimes, this model exhibits spin-triplet superconductivity\,\cite{lin1986condensation,penc1994phase,lin1995phase,xiang2019doping,shinjo2019machine,qu2022spin}. Its extension to two dimensions (2D) further reveals robust spin-triplet pairing instability near quarter filling\,\cite{onari2004phase,huang2013unconventional,nayak2018exotic,van2018extended,chen2023superconducting,cao2025dominant,farid2025lifshitz}. Although once treated as an artificial toy model, mixed-sign interactions have been experimentally identified in strongly correlated cuprates in quasi-1D thin films and single crystals\,\cite{chen2021anomalously, padma2025beyond, scheie2025cooper, Li2024Doping}. In these systems, phonons are believed to mediate attractive interactions between nearest-neighbor electrons while preserving dominant on-site repulsion, leading to a scenario effectively described by the EHM\,\cite{wang2021phonon,tang2023traces,feiguin2023effective,shen2024signatures,tang2024influence}. Although further experimental validation is needed to quantify these interactions, such mechanisms may be broadly realized in Ruddlesden-Popper transition-metal oxides due to structural similarities, providing a promising route toward stabilizing spin-triplet pairing beyond quasi-1D materials.

\begin{figure}[!b]
\begin{center}
\includegraphics[width=8.5cm,clip]{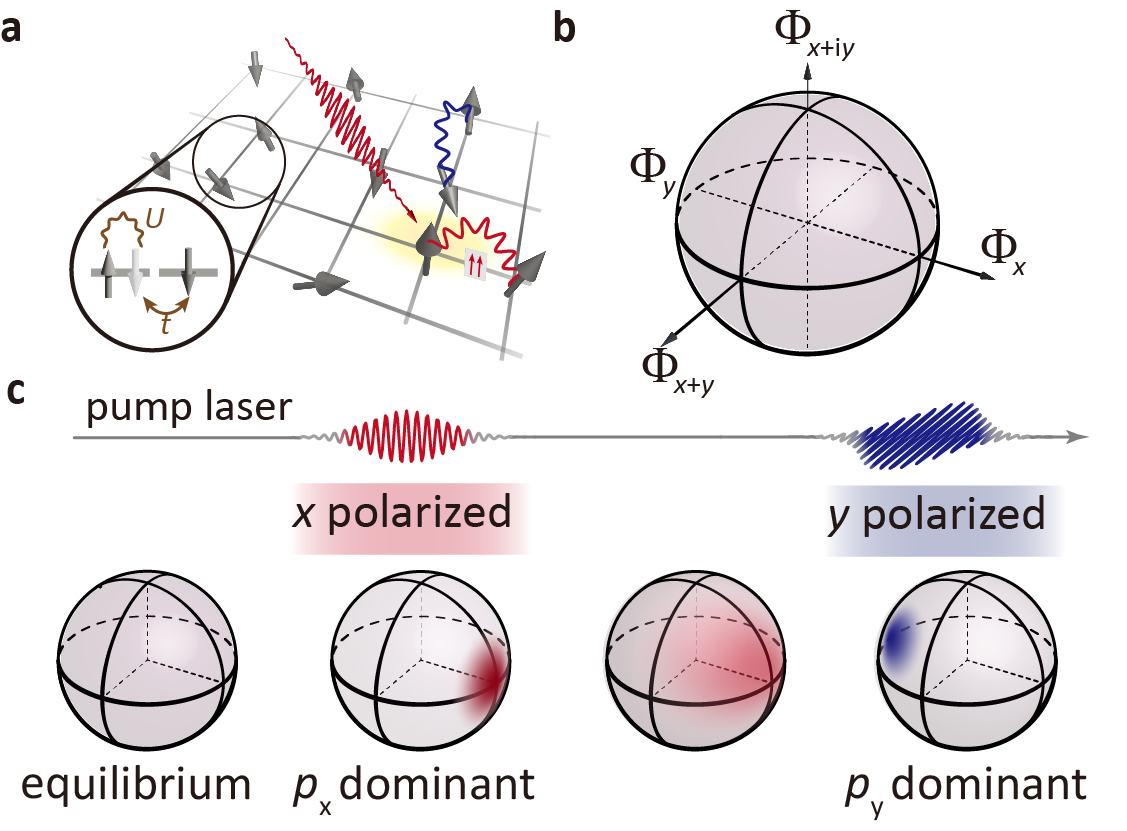}\vspace{-3mm}
\caption{\label{fig:cartoon}
\textbf{Schematic of light-controlled spin-triplet pairing.} \textbf{a} Correlated electrons on a 2D lattice, characterized by hopping $t$, on-site Coulomb interaction $U$, and nearest-neighbor interaction $V$. An external pump pulse transiently induces a ferromagnetic spin-exchange interaction along the pump polarization direction. \textbf{b} Bloch sphere representation of pairing correlations. \textbf{c} Schematic illustrating the redistribution of the pairing correlations across the Bloch sphere, following an optical pump and a subsequent second pump.
}
\end{center}
\end{figure}

Another challenge in harnessing spin-triplet superconductivity is achieving efficient and selective control over its order parameter. Recent time-dependent mean-field studies have proposed sophisticated strategies for manipulating multiple $p$-wave order parameters\,\cite{claassen2019universal}, and inducing transitions between spin-singlet and triplet states\,\cite{gassner2024light}. In mean-field approximations, where quantum fluctuations are neglected, these approaches typically require carefully timed sequences of pump pulses with varying polarizations. In strongly correlated systems, light-induced superconductivity may be more directly accessible due to the interplay of competing orders and quantum fluctuations. Experimental evidences from cuprates and molecular crystals suggest that nonequilibrium optical excitation can dynamically alter superconducting states in correlated materials\,\cite{fausti2011light,nicoletti2014optically,hu2014optically,mankowsky2014nonlinear,  buzzi2020photomolecular,buzzi2021phase}.

Building on these developments, we propose a dynamical strategy to engineer spin-triplet \rvss{pairing} in strongly correlated systems with mixed-sign interactions. Specifically, we leverage the Floquet engineering of magnetic exchange interactions\,\cite{mentink2014ultrafast,mentink2015ultrafast} to analysis how a transient sign reversal of the exchange interaction along the pump polarization,  when the pump laser is tuned near resonance [see Fig.~\ref{fig:cartoon}]. Using time-dependent exact diagonalization simulations on a extended Hubbard model, we demonstrate the feasibility of selectively enhancing existing spin-triplet pairing correlations on ultrafast timescales. A subsequent pulse with an orthogonal polarization can further manipulate the order parameter, enabling controlled switching between pairing symmetries. This approach paves the way for designing and dynamically controlling spin-triplet superconductivity in strongly correlated materials.

\section{Results}
\subsection{Light-Enhanced Spin-Triplet Pairing Correlations}

To investigate spin-triplet pairing instabilities in strongly correlated systems, we focus on a quarter-filled (\textit{i.e.}, 50\% hole-doped) EHM on a square lattice, which serves as a prototypical model for transition-metal oxides. Following Eq.~\eqref{eq:EHM} in \textbf{Methods}, the EHM incorporates a repulsive on-site interaction $U$ and an attractive nearest-neighbor interaction $V$. Previous studies have demonstrated that such a model supports strong spin-triplet pairing instabilities\,\cite{lin1986condensation,penc1994phase, lin1995phase,xiang2019doping, shinjo2019machine, onari2004phase, huang2013unconventional, nayak2018exotic,van2018extended}, confirmed through many-body numerical simulations\,\cite{qu2022spin,chen2023superconducting,cao2025dominant,farid2025lifshitz}. Here, we adopt $U=8t_h$ and $V=-t_h$ as model parameters in Eq.~\eqref{eq:EHM}, consistent with experimental estimates for cuprates\,\cite{chen2021anomalously, padma2025beyond, scheie2025cooper}.

The spin-triplet pairing instabilities are characterized by the three-fold pairing operators for each spatial symmetry. For example, the $p_x$-wave pairing operator is: 
\begin{equation}
    \begin{aligned}
        \Delta^{p_x}_{\mathbf{i},0} &= \frac{1}{\sqrt{2}} \left(c_{\mathbf{i},\uparrow} c_{\mathbf{i}+\hat{x},\downarrow} + c_{\mathbf{i},\downarrow} c_{\mathbf{i}+\hat{x},\uparrow}\right)\,,\\
    \Delta^{p_x}_{\mathbf{i},1} &=  c_{\mathbf{i},\uparrow} c_{\mathbf{i}+\hat{x},\uparrow}\,,\quad
    \Delta^{p_x}_{\mathbf{i},-1} =  c_{\mathbf{i},\downarrow} c_{\mathbf{i}+\hat{x},\downarrow}\,,
\end{aligned}
\end{equation}
where $c_{\mathbf
{i},\sigma}$ represents the annihilation operator for an electron with spin $\sigma$ at site $\mathbf{i}$. Analogous operators for the $p_y$, $p_{x}\pm p_{y}$, and $p_x\pm ip_{y}$ pairing channels are defined in the Supplementary Note 1. For any many-body state $|\psi\rangle$, the $p$-wave pairing correlations are computed as:
\begin{eqnarray}\label{eq:pairingCorr}
\Phi_\alpha=\sum_{\mathbf{i},\mathbf{j}}\langle \psi | \Delta^{p_\alpha ^\dagger}_\mathbf{i}\Delta^{p_\alpha}_\mathbf{j} | \psi\rangle,
\end{eqnarray}
where $\alpha$ labels the pairing symmetry and the pairing operator is averaged over the triplet states $\Delta^{p_\alpha}_\mathbf{i} = (\Delta^{p_\alpha}_{\mathbf{i},1}+\Delta^{p_\alpha}_{\mathbf{i},0}+\Delta^{p_\alpha}_{\mathbf{i},-1})/\sqrt{3}$. These pairing correlations can be represented as vectors on the Bloch sphere (see Fig.~\ref{fig:cartoon}\textbf{b}), providing a geometric perspective on the interplay between different pairing symmetries. In this work, we focus on analyzing \rvss{short-range} pairing correlations within finite clusters rather than long-range order, the latter of which remains challenging to establish in 2D systems due to strong quantum fluctuations.

\begin{figure*}[!t]
\begin{center}

\includegraphics[width=17cm,clip]{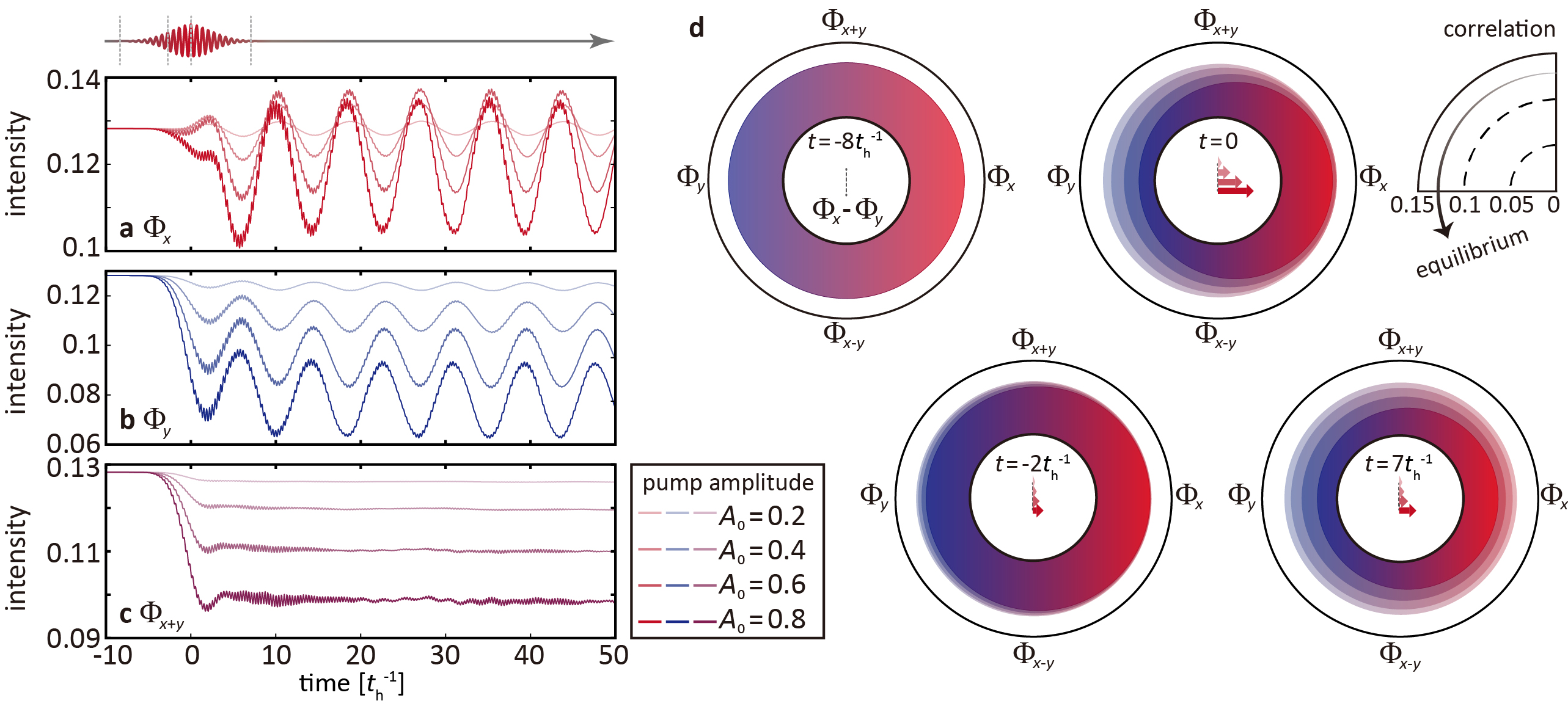}\vspace{-5mm}
\caption{\label{fig:singlePumpDynamics}
\textbf{Time evolution of spin-triplet pairing correlations under a single pump.} \textbf{a}-\textbf{c} Dynamics of pairing correlations in the \textbf{a} $p_x$, \textbf{b} $p_y$, and \textbf{c} $p_x+p_y$ channels, in response to a $\Omega=9.2t_h$ pump pulse with amplitudes $A_0=0.2$, 0.4, 0.6 and 0.8, respectively. The top inset displays the time-dependent vector potential of the pump field. \textbf{d} Distribution of pairing correlations projected onto the equatorial plane of the Bloch sphere at four representative time points (marked by gray dashed lines in the inset of \textbf{a}). Colors from light to dark indicate increasing pump strengths from $A_0=0.2$ to 0.8. The arrows at the center indicate the magnitude of $\Phi_x -\Phi_y$ for each pump strength.
}
\end{center}
\end{figure*}

At equilibrium, the $p$-wave pairing correlations, which are isotropic across all six orientations, exhibit a pronounced hump for the interaction parameters adopted in this study ($U=8t_h$ and $V=-t_h$), signaling strong spin-triplet pairing instabilities\,\cite{chen2023superconducting}. Starting from this equilibrium state, we investigate the nonequilibrium dynamics of $p$-wave pairing correlations induced by an external optical excitation. The time-dependent wavefunction $|\psi(t)\rangle$ is obtained by evolving the system under a time-dependent Hamiltonian, where the external vector potential $\mathbf{A}(t)$ is incorporated via the Peierls substitution (see \textbf{Methods}). The instantaneous pairing correlation $\Phi_\alpha(t)$ is then calculated by substituting $|\psi(t)\rangle$ into Eq.~\eqref{eq:pairingCorr}. For the single-pump setup considered in this section, the external light pulse is usually modeled as a Gaussian envelope: 
\begin{equation}\label{eq:singlePulse}
    \mathbf{A}(t) = A_0\hat{\textbf{e}}_{\rm pol}e^{-t^2/2\sigma_t^2}\cos(\Omega t+\phi)\,, 
\end{equation}
where $A_0$, $\Omega$, and $\hat{\textbf{e}}_{\rm pol}$ represent the pump amplitude, frequency, and polarization, respectively.

Without loss of generality, we focus on a single $x$-polarized laser pulse, $\hat{\textbf{e}}_{\rm pol} = \hat{x}$, to explore the anisotropic response of pairing correlations\,\cite{polarizationSymmetry}. In typical pump-probe experiments, phase coherence between the pump and probe lasers is often not strictly maintained, leading to phase-averaged measurements that suppress phase-locked oscillations. To account for this, we evaluate the phase-averaged pairing correlation as $\sum_n \Phi_\alpha(t; \phi = \phi_n)/n$, with a detailed analysis provided in the Supplementary Note 2. For clarity, all figures in the main text present results with $\phi = 0$, as the phase slightly influences the amplitude of $\Phi_\alpha(t)$ without qualitatively altering its overall temporal evolution.

The dynamics of spin-triplet pairing correlations under an $x$-polarized laser pulse with $\Omega = 9.2 t_h$ are presented in Figs.~\ref{fig:singlePumpDynamics}\textbf{a}-\textbf{c}. A prominent feature of this response is the pronounced oscillatory behavior observed in $\Phi_x(t)$ and $\Phi_y(t)$, which exhibit a striking anti-phase relationship both during and after the pump pulse. This anti-phase behavior reflects an intrinsic competition between these orthogonal pairing channels. While increasing the pump strength enhances the oscillation amplitude, the underlying anti-phase structure remains unchanged, indicating the robustness of this dynamical competition. As a result, the pairing channel aligned with the pump field (\textit{e.g.}, $p_x$) is transiently enhanced at specific oscillation phases, such as near $t=10t_h^{-1}$, at the cost of significantly suppressing its orthogonal counterpart (\textit{e.g.}, $p_y$). 

In contrast, other orthogonal pairs of pairing channels, such as $\Phi_{x\pm y}(t)$, exhibit a different dynamical response [see Fig.~\ref{fig:singlePumpDynamics}\textbf{c}]. These channels undergo an abrupt suppression during the pump pulse and show minimal oscillatory behavior.  Instead, their post-pump evolution reflects an averaged behavior of $\Phi_x(t)$ and $\Phi_y(t)$, suggesting a net suppression of pairing correlations, likely driven by photo-induced carrier screening effects\,\cite{wang2021fluctuating}. The maximum enhancement of $\Phi_x(t)$ follows a non-monotonic dependence on pump strength, dictated by the competing effects of light-enhanced $p_x$ pairing and the overall suppression of pairing correlations.

To further elucidate the anisotropic response of laser-induced $p$-wave pairing correlations, we map the pairing correlations onto the Bloch sphere at four representative time points: $t=-8t_h^{-1}$
, $-2t_h^{-1}$, $0$ and $7t_h^{-1}$ [see Fig.~\ref{fig:singlePumpDynamics}\textbf{d}]. At equilibrium, pairing correlations are isotropic without any symmetry breaking. However, once a linearly polarized pump is applied, this symmetry is broken, causing the Bloch sphere to shift preferentially toward the $p_x$ direction while suppressing the $p_y$ component. This asymmetry becomes more pronounced with higher pump fluence, indicating a stronger laser-induced anisotropy. After the pump, the Bloch sphere undergoes continuous oscillations along the horizontal axis, yet $\Phi_x(t)$ consistently dominates over $\Phi_y(t)$. Notably, the center of the Bloch sphere does not cross the origin, as shown by the directional arrows in Fig.~\ref{fig:singlePumpDynamics}\textbf{d}, reflecting sustained $p_x$ channel dominance triggered by the pump. The first maximal imbalance occurs near $t \sim \sigma = 2.5t_h^{-1}$, with slight variations depending on the pump amplitude. Since the correlated system is closed and lacks dissipation, the undamped laser energy drives continuous Bloch sphere oscillations over extended timescales lie within the pre-thermal plateau\,\cite{aoki2014nonequilibrium,de2021colloquium}, ensuring the persistence of the $p_x$ dominance and enhancement of spin-triplet pairing after the pump pulse ends\,\cite{kalthoff2019floquet,sentef2020quantum}.

\subsection{Floquet Engineering of Anisotropic Spin Interactions}\label{sec:freqDep}

\begin{figure}[!t]
\begin{center}
\includegraphics[width=8.5cm,clip]{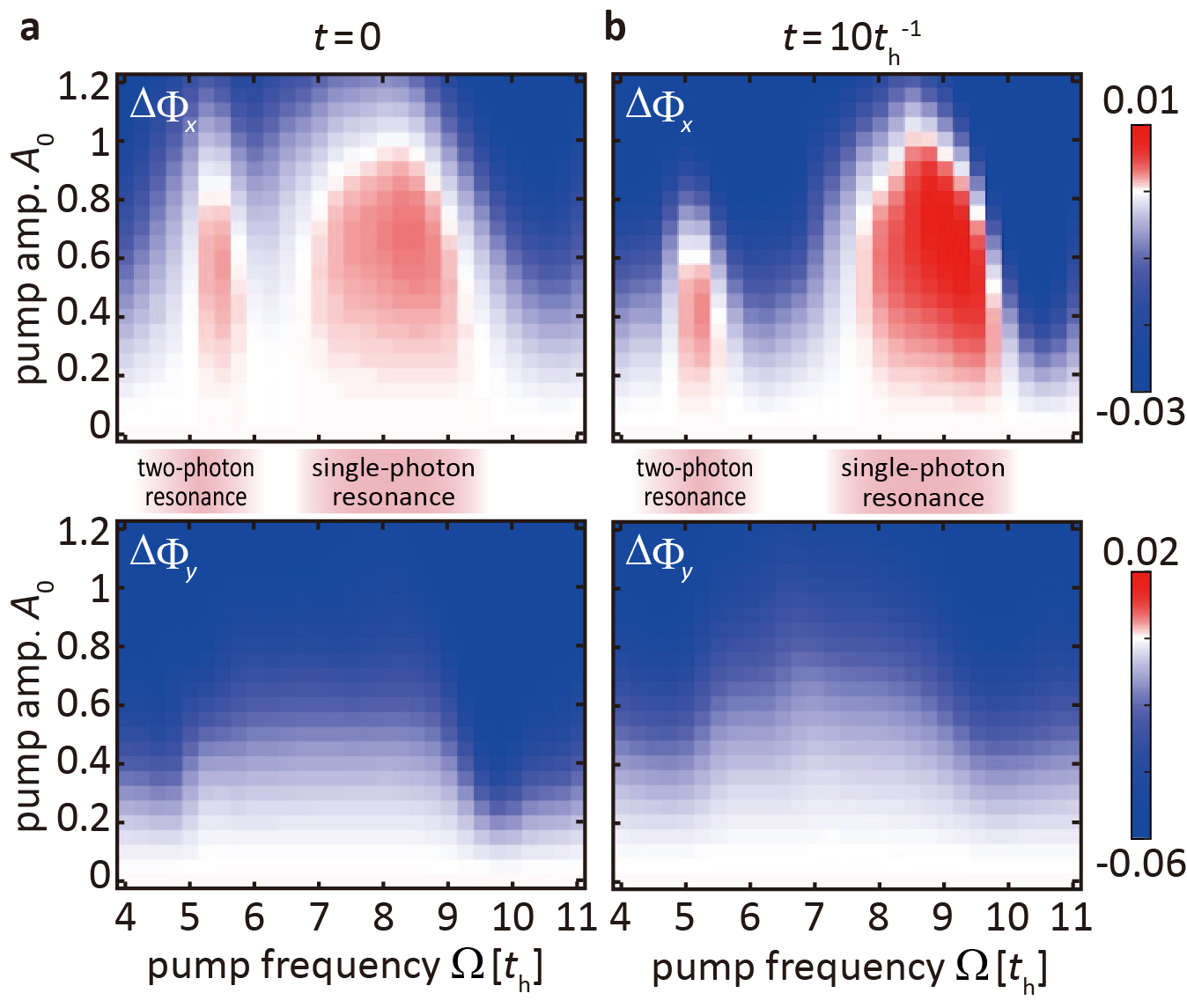}\vspace{-3mm}
\caption{\label{fig:frequencyDependence}
\textbf{Pump frequency and amplitude dependence.} \textbf{a} Pump-induced changes in $p_x$ (upper) and $p_y$ (lower) pairing correlations at $t=0$ (center of the pump field), relative to their equilibrium values, plotted as a function of pump frequency and amplitude. The two distinct regimes of $p_x$ enhancement correspond to the single- and two-photon resonances. \textbf{b} Same as \textbf{a} but evaluated at $t = 10 t_h^{-1}$. }
\end{center}
\end{figure}

The selective enhancement of spin-triplet pairing correlations reflects an underlying control mechanism mediated by anisotropic spin interactions. To examine this mechanism, we simulate the nonequilibrium dynamics of $\Phi_x(t)$ and $\Phi_y(t)$ across a broad range of pump amplitudes and frequencies. To isolate light-induced modifications, we focus on the differential pairing correlation as $\Delta\Phi_\alpha(t) = \Phi_\alpha(t) - \Phi_\alpha(t=-\infty)$ by offsetting with their equilibrium values. Fig.~\ref{fig:frequencyDependence}\textbf{a} presents $\Delta\Phi_x$ and $\Delta\Phi_y$ at the pump pulse center ($t = 0$). Although increasing the pump amplitude generally suppresses the pairing correlations across all channels due to photocarrier screening, a pronounced enhancement of $\Phi_x$ emerges near $\Omega \approx U = 8t_h$, persisting across all fluences. We further investigate light-induced changes in $\Delta\Phi_x(t)$ and $\Delta\Phi_y(t)$ at $t=10 t_h^{-1}$ (see Fig.~\ref{fig:frequencyDependence}\textbf{b}), corresponding to the first anti-phase oscillation peak observed in Fig.~\ref{fig:singlePumpDynamics}. In both time points, two distinct frequency regimes emerge: a primary enhancement peak near $\Omega \sim 9t_h = U-V$, consistent with our earlier findings in Fig.~\ref{fig:singlePumpDynamics}, and a secondary, narrower peak around $\Omega \sim 5\,t_h$. In contrast, $\Delta\Phi_y(t)$ is suppressed in all pump conditions.

\begin{figure*}[!t]
\begin{center}
\includegraphics[width=17cm,clip]{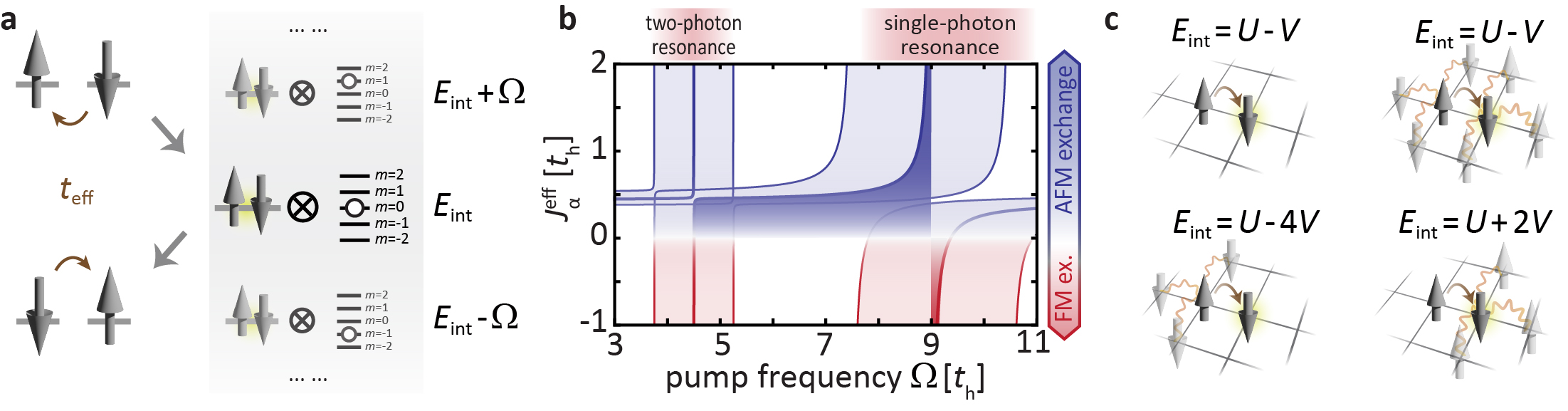}\vspace{-3mm}
\caption{\label{fig:FloquetIllustration} \textbf{Floquet-engineered anisotropic spin exchange.}
\textbf{a} Schematic depiction of the spin exchange process in a Floquet system. The effective superexchange interaction $J$ originates from second-order virtual hopping processes that involve multiple intermediate states, with their energies set by the equilibrium intermediate-state energy $E_{\text{int}}$ and integer multiples of the pump frequency $\Omega$. \textbf{b} Effective spin exchange interaction $J_{\alpha}^{\rm eff}$ at a transient state, calculated using the Floquet theory, as a function of the pump frequency $\Omega$. The red and blue regions indicate ferromagnetism (FM) and antiferromagnetism (AFM) exchange interactions, respectively. The solid lines correspond to calculations assuming homogeneous electron occupations, while the translucent extended $J_{\alpha}^{\rm eff}$ resonance area due to occupation imbalances, as illustrated in \textbf{c}. The single- and two-photon resonances are determined by the variation of FM regions. \textbf{c} Schematic illustration of various intermediate state configurations with distinct electron occupations and their corresponding intermediate-state energies $E_{\text{int}}$.
}
\end{center}
\end{figure*}

The observed frequency-dependent enhancement of spin-triplet pairing correlations in the dynamical simulation of the EHM model can be attributed to Floquet engineering of effective spin-exchange interactions. In the presence of a periodic pump field, the transient system can be approximated by an effective Floquet Hamiltonian, treating the laser field as periodic\,\cite{oka2019floquet,rudner2020band,de2021colloquium}. Within this framework, the monochromatic light field couples to the electronic states, creating a series of photon-dressed states with energy levels spaced by multiples of the pump frequency $\Omega$\,\cite{eckardt2015high}. In strongly correlated systems, such photon dressing significantly modifies the intermediate states involved in spin superexchange process\,\cite{mentink2014ultrafast,mentink2015ultrafast}. As illustrated in Fig.~\ref{fig:FloquetIllustration}\textbf{a}, an intermediate state with double occupancy transforms into photon-dressed states under periodic driving, effectively renormalizing the spin superexchange interactions. Consequently, the Floquet-engineered spin exchange interaction becomes\,\cite{mentink2014ultrafast,mentink2015ultrafast}:
\begin{eqnarray}\label{eq:FloquetJ}
J_{\alpha}^{\rm eff}(\mathbf{A}, \Omega) = \sum_{m=-\infty}^{\infty} \frac{4t_{h}^2\mathcal{J}_m(A_0\cdot \hat{\mathbf{e}}_{\alpha})^2}{E_{\rm int}+m\Omega}\,,
\end{eqnarray}
where $\mathcal{J}_m(x)$ is the $m$th-order Bessel function of the first kind, describing contributions from processes involving $m$ photons with energy $\Omega$. Here, $E_{\rm int}$ represents the energy cost of creating a doubly occupied intermediate state relative to the singlet ground state. Crucially, the effective exchange interaction $J_{\alpha}^{\mathrm{eff}}$ depends on the projection $A_0\cdot\hat{\mathbf{e}}_{\alpha}$ of the pump field along the exchange pathway, allowing dynamic control over both the magnitude and the sign of the interaction by tuning the pump polarization, amplitude, and frequency\,\cite{wang2021x}.

In contrast to the Hubbard model, where the intermediate-state energy is determined by $U$ alone\,\cite{mentink2014ultrafast, mentink2015ultrafast}, the EHM with on-site $U$ and nearest-neighbor $V$ results in an intermediate-state energy $E_{\mathrm{int}} \approx U - V$, setting the single-photon resonance at $\Omega \approx 9t_h$ in Eq.~\eqref{eq:FloquetJ}. Fig.~\ref{fig:FloquetIllustration}\textbf{b} presents the simulated spin exchange for $A_0=0.4$ with $m$ truncated to $\pm50$, where $J_{x}^{\rm eff}$ exhibits strong pump frequency dependence. Away from resonance, the Floquet-engineered effective $J_{x}^{\rm eff}$ recovers its equilibrium antiferromagnetic (AFM) exchange interaction $J\sim 4t_h^2/E_{\mathrm{int}}\sim0.5t_h$. However, as $\Omega$ approaches resonance with $E_{\mathrm{int}}$, $J_{x}^{\rm eff}$ undergoes significant renormalization. Notably, when $\Omega$ slightly exceeds $E_{\rm int}$, the exchange interaction undergoes a sign reversal, transitioning from AFM to ferromagnetic (FM) (red-shaded region). This sign flip is crucial because AFM exchange favors spin-singlet pairing, while FM exchange supports spin-triplet pairing, thereby explaining the enhancement of $\Phi_x$ for $\Omega \gtrsim E_{\mathrm{int}}$. Moreover, the Floquet effect is polarization-selective: the exchange interaction along the $y$ direction ($J_{y}^{\rm eff}$) remains largely unaffected due to its orthogonality with the pump polarization, leading to the observed anisotropic pairing correlations.

The resonance condition for Floquet-induced FM exchange is not confined to the narrow frequency window denoted by the red-shaded regime in Fig.~\ref{fig:FloquetIllustration}\textbf{b}, but extends across a broader range of pump frequencies. This broadening arises from two primary effects. First, the above simplified Floquet analysis, which is based on atomic models, requires adjustments to incorporate band dispersion and quantum fluctuations in many-body systems. Second, the intermediate-state energy $E_{\mathrm{int}}$, which determines $J_x^{\rm eff}$ in Eq.~\eqref{eq:FloquetJ}, is not a fixed quantity in the doped EHM but varies across lattice sites in a given Fock state, depending on the local occupation configurations. As illustrated in the upper panels of Fig.~\ref{fig:FloquetIllustration}\textbf{c}, configurations with isolated spin pairs or uniform neighbor densities yield $E_{\rm int}=U-V$, corresponding to a strong resonance near $\Omega \sim U-V = 9t_h$ (solid lines in Fig.~\ref{fig:FloquetIllustration}\textbf{b}). However, non-uniform occupations introduce a distribution of $E_{\rm int}$ values spanning from  $U-4V$ to $U+2V$. At an average filling of $\langle n \rangle = 0.5$, this variation extends the resonance window from $U+0.5V=7.5t_h$ to $U-2.5V = 10.5t_h$, as indicated by translucence red and blue area in Fig.~\ref{fig:FloquetIllustration}\textbf{b}. This asymmetric broadened resonance range around the center $\Omega = U-V$ aligns with the observed light-enhanced $p_x$ pairing instabilities in Fig.~\ref{fig:frequencyDependence}, reflecting the role of Floquet engineering in tuning pairing correlations. 

Beyond the dominant single-photon resonance, Floquet theory also predicts a two-photon resonance condition, occurring when the driving frequency satisfies $\Omega \approx E_{\mathrm{int}}/2 = 4.5\,t_h$. Variations in local occupations further modulate $E_{\mathrm{int}}$, yielding a relatively narrower yet finite resonant window for the two-photon process, spanning from $3.75t_h$ to $5.25t_h$. The frequency dependence observed in the simulated $\Phi_x(t)$ (Fig.~\ref{fig:frequencyDependence}) also aligns well with this prediction, further corroborating the role of Floquet engineering in modulating pairing correlations. Additional details on the pairing correlation dynamics at $\Omega = 5.2t_h$ are provided in Supplementary Note 3. 

Beyond the Floquet-engineered spin exchange $J_x^{\rm eff}$, the laser field also modifies the electronic hopping amplitude as $t_x^{\rm eff}\sim t \mathcal{J}_0(A_0\cdot \hat{\mathbf{e}}_{\alpha})$, thereby inducing anisotropy between the $x$ and $y$ directions. Although this anisotropy resembles the effect of uniaxial strain and does contribute to the spin-triplet pairing,\cite{gassner2024light}, it does not generate the mixed AFM–FM exchange interactions responsible for the resonances observed in Fig.\ref{fig:frequencyDependence}, and thus does not serve as the primary mechanism behind the light-induced superconductivity in the EHM. Nevertheless, the suppression of $t_x^{\rm eff}$ constrains the strength of light-enhanced pairing, which explains the observed saturation and eventual decline of $\Phi_x$ around $A_0 \sim 1$ in Figs.~\ref{fig:frequencyDependence}\textbf{a} and \textbf{b}.

\subsection{Ultrafast Switching Between Different Channels}\label{sec:switch}

The ability to selectively enhance $p$-wave superconductivity along a given direction raises a compelling question: can optical excitation be used to switch the dominant pairing channel to an orthogonal one? Within a mean-field framework, transitions between different pairing channels like $p_x$ and $p_y$ (or between chiral states $p_x+ip_y$ and $p_x-ip_y$) are inherently challenging due to the strict orthogonality of their order parameters. Typically, such a switch requires a carefully designed sequence of linear and circular pulses to navigate a path on the Bloch sphere that connects the two poles\,\cite{claassen2019universal,yu2021optical}. However, in strongly correlated many-body systems, quantum fluctuations naturally broaden the distribution of pairing correlations across the Bloch sphere rather than confining them to discrete symmetry-breaking points. This broader distribution may allow for more flexible control pathways, enabling direct switching between orthogonal $p$-wave channels without requiring finely tuned orchestrated pulse sequences.

To test this hypothesis, we simulate the response of spin-triplet pairing correlations to two sequential linear pulses with orthogonal polarizations. The first $x$-polarized pulse serves to establish light-induced $p_x$ pairing correlations, as previously discussed, while the second pulse is designed specifically to validate the switching mechanism. Thus, the vector potential from Eq.~\eqref{eq:singlePulse} is modified as follows: 
\begin{equation}\label{eq:twoPulses}
\begin{aligned}
    \mathbf{A}(t) =\,& A_0\hat{\textbf{e}}_{\rm x}e^{-t^2/2\sigma_t^2}\cos(\Omega t+\phi)\\
    &+A_0^\prime\hat{\textbf{e}}_{\rm y}e^{-(t-\Delta t)^2/2\sigma_t^2}\cos[\Omega (t-\Delta t)+\phi^\prime]\,, 
\end{aligned}
\end{equation}
where $\Delta t$ denotes the time delay between the two pump pulses and the relative phase $\phi'-\phi$ controls their coherence. Here, we set both $\phi$ and $\phi'$ to zero, while the impact of phase-averaging is discussed later.

\begin{figure}[!t]
\begin{center}
\includegraphics[width=8.5cm]{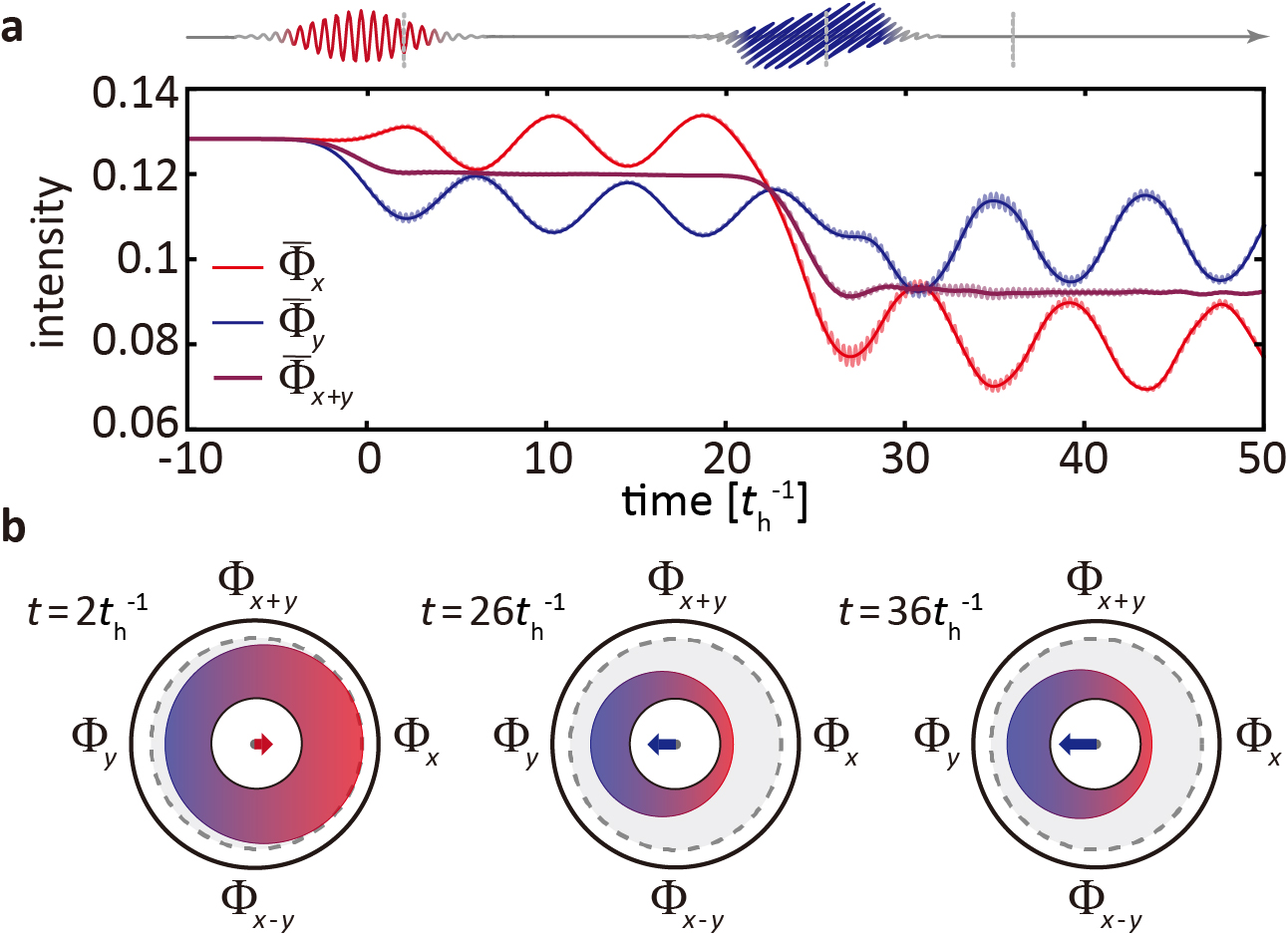}\vspace{-3mm}
\caption{
\textbf{Ultrafast switch between pairing correlations.} \textbf{a} Dynamics of the $p_x$, $p_y$, and $p_x+p_y$ pairing correlations induced by two sequential pulses with $x$ and $y$ polarizations. Solid lines represent the phase-averaged pairing correlation  $\overline{\Phi}_\alpha$, while the translucent lines correspond to results with a fixed phase $\phi = 0$. The upper inset shows the vector potential of the pump field. \textbf{b} Redistribution of pairing correlations on the equatorial plane of the Bloch sphere at the three selected time points (marked in \textbf{a}). The central arrows indicate the relative magnitude of $\Phi_x - \Phi_y$ at the corresponding time.}
\label{fig:doublePump}
\end{center}
\end{figure}

Figure \ref{fig:doublePump}\textbf{a} presents the evolution of spin-triplet pairing correlations under the influence of two sequential laser pulses. The first laser pulse, $x$-polarized with an amplitude $A_0 = 0.4$ and frequency $\Omega = 9.2\,t_h$ (same as Fig.~\ref{fig:singlePumpDynamics}), triggers the symmetry breaking and enhances the $p_x$ pairing correlations. After several oscillation cycles, significantly exceeding the pulse duration, a second $y$-polarized laser pulse is applied at a time delay $\Delta t=25\,t_h^{-1}$, maintaining the same pump frequency and phase. To ensure an active switching of the dominant pairing channel, rather than merely canceling the pre-established correlations (see Supplementary Note 4 for detailed discussions), the second pulse amplitude is set to $A_0^\prime=2A_0$ (see Fig.~\ref{fig:doublePump}\textbf{a}). Remarkably, this second pulse rapidly reverses the dominance between $\Phi_x$ and $\Phi_y$. The $p_x$ pairing correlation experiences a substantial suppression, with its average intensity reduced by roughly 43\%, while the $p_y$ pairing correlation remains relatively less affected, due to the interplay between light-enhanced pairing effects and additional photo-induced carriers.

The dynamics underlying this ultrafast
 switching are best visualized through the Bloch sphere representation. In contrast to mean-field dynamics, where the order parameter is represented as a single point on the Bloch sphere, many-body pairing correlations $\Phi_\alpha$ are distributed over the sphere. As shown in Fig.~\ref{fig:doublePump}\textbf{b}, the $y$-polarized pulse triggers a redistribution of these correlations, effectively shifting the spectral weight towards the $\Phi_y$ direction while depleting the previously dominant $\Phi_x$ component stabilized by the initial pulse. This collective motion results in a persistent enhancement of $\Phi_y$ relative to $\Phi_x$, with the centroid of the distribution clearly shifting towards the $\Phi_y$ along the $\Phi_x-\Phi_y$ axis. For the chosen pulse strength $A_0^\prime = 2A_0$, $\Phi_y$ consistently dominates after the second pulse (see the arrows in Fig.~\ref{fig:doublePump}\textbf{b}), confirming the ultrafast switching mechanism driven solely by a single orthogonal pulse.

By varying the phases $\phi$ and $\phi^\prime$, we further explore the influence of coherence on the ultrafast switch. In Fig.~\ref{fig:doublePump}\textbf{a}, the thick lines show the phase-averaged values of $\overline{\Phi}_x$, $\overline{\Phi}_y$, and $\overline{\Phi}_{x+y}$, computed by averaging over 20 different phases while keeping all other parameters fixed. Our results indicate that while minor high-frequency oscillations exhibit phase sensitivity, the overall envelope closely follows the previously discussed phase-locked behavior. This insensitivity arises from the pairing correlation in Eq.~\eqref{eq:pairingCorr}, which effectively eliminates the gauge degrees of freedom. As a result, both the light-induced enhancement of spin-triplet pairing and the ultrafast switching between pairing channels remain robust against variations in pump pulse coherence.

\section{Discussion}
Because of the need to preserve $C_4$ symmetry and the inherent numerical difficulties of simulating nonequilibrium dynamics, our analysis is based on exact diagonalization of correlated electrons within a finite cluster, where spontaneous symmetry breaking cannot occur. Accordingly, we focus on short-range pairing correlations rather than off-diagonal long-range order, whose identification would require extrapolation to the 2D thermodynamic limit --- a task currently beyond the reach of existing many-body computational techniques. Despite this limitation, short-range quantum fluctuations are a ubiquitous feature of strongly correlated systems. At equilibrium, fluctuation-driven non-symmetry-breaking gaps have been experimentally observed in unconventional superconductors, such as cuprates\,\cite{kondo2015point,he2021superconducting} and FeSe\,\cite{faeth2021incoherent,xu2021spectroscopic}, as well as in other phase transitions\,\cite{schafer2001high,chatterjee2015emergence,chen2023role}. These fluctuations become even more pronounced in low-dimensional or nanoscale materials. More importantly, light-induced superconductivity exhibits even stronger quantum fluctuations with shorter-range correlations\,\cite{boschini2018collapse,sun2020transient,wang2021fluctuating}. These short-range correlations manifest as superconducting gaps with limited coherence, making them effectively indistinguishable from long-range orders in pump-probe spectroscopic measurements.

In strongly correlated systems, the inherent flexibility of pairing correlations offers significant advantages for laser-based control compared, beyond the constraints of mean-field order parameters. As demonstrated above, spin-triplet pairing correlations can be effectively engineered through Floquet manipulation of spin-exchange interactions $J$ along the pump polarization. Furthermore, the distributed nature of these correlations across the Bloch sphere enables a straightforward single-pulse strategy to switch between orthogonal pairing channels, which would otherwise necessitate more sophisticated control protocols in a mean-field framework\,\cite{claassen2019universal,yu2021optical}, or rely on static uniaxial strain or Jahn–Teller distortions\,\cite{gassner2024light}. These fundamental differences suggest that many-body quantum fluctuations, rather than acting as a limiting factor, can instead serve as a powerful resource for the dynamic control of spin-triplet superconducting states. Such a distinction may also account for the fact that light-induced superconductivity has been frequently observed in correlated materials\,\cite{fausti2011light,nicoletti2014optically,hu2014optically,mankowsky2014nonlinear, buzzi2020photomolecular,buzzi2021phase}.

Our study focuses on the ultrafast regime soon after the pump finishes. Based on estimated hopping parameters typical of cuprates (see Methods), the time interval shown in Figs.~\ref{fig:singlePumpDynamics} and \ref{fig:doublePump} spans approximately $\sim 100$\,fs. This short interval lies within the prethermal regime, as identified in related systems\,\cite{aoki2014nonequilibrium,de2021colloquium}. If the simulation time were extended substantially, it would be necessary to account for additional effects such as heating, decoherence, and the retarded response of phonons.

\section{Methods}
\subsection{Extended Hubbard Model}

The extended Hubbard model (EHM) and its variants effectively describe correlated electrons in transition-metal oxides. The Hamiltonian is defined as
\begin{equation}\label{eq:EHM}
\mathcal{H} = -t_h \sum_{\langle ij \rangle, \sigma} (c^\dagger_{i\sigma} c_{
j\sigma} + h.c.)
+ U \sum_i n_{i\uparrow} n_{i\downarrow} + V\sum_{\langle ij \rangle} n_i n_j.
\end{equation}
where $c_{i\sigma} (c^\dagger_{i\sigma})$ annihilates (creates) an electron at site $i$ with spin $\sigma=\uparrow,\downarrow$ and $n_{i\sigma} = c^\dagger_{i\sigma} c_{i\sigma} (n_i = n_{i\uparrow}+n_{i\downarrow})$ represents the electron density operator. The model incorporates three key parameters: the nearest-neighbor hopping amplitude $t_h$, the on-site Coulomb repulsion $U$, and the nearest-neighbor interaction $V$. In this study, the on-site $U$ is always repulsive, accounting for the double-occupation energy, while the nearest-neighbor interaction $V$ is attractive, representing an effective phonon-mediated interaction. We adopt model parameters $U=8t_h$ and $V=-t_h$, based on prior experimental estimates for 1D cuprate chains\,\cite{chen2021anomalously, wang2021phonon}, while the conclusion broadly applies to a variety of model parameters (see Supplementary Note 5). The simulation is performed within a canonical ensemble framework, where the total particle number is fixed to ensure quarter filling, and the temperature is set to absolute zero. For $t_h=300-400\,$meV, a 400-nm laser resides in the single-photon resonance and an 800-nm laser resides in the two-photon resonance. Simulations are conducted on a 4$\times$4 cluster with periodic boundary conditions.

\subsection{Peierls substitution and the Krylov-Subspace Method}

To describe the light-driven dynamics, we incorporate the Peierls substitution $t_hc^\dagger_{i\sigma}c_{j\sigma} \xrightarrow[]{} t_he^{i\int_{r_i}^{r_j}{\textbf{A}(t)d\textbf{r}}}c^\dagger_{i\sigma}c_{j\sigma}$, which serves as an effective approximation of the light-matter interaction within the second-quantization framework. The specific pump vector potential $\textbf{A}(t)$ is provided by Eqs.~\eqref{eq:singlePulse} and \eqref{eq:twoPulses} of the main text. We employ the SLEPc library to solve the ground state\,\,\cite{hernandez2005slepc}, which yields a three-fold degenerate ground state due to the hidden hypercubic symmetry of the $4\times4$ cluster. The expectation value of pairing correlations $\Phi_\alpha$ is taken as an average over all degenerate states and all triplet $S_z$ states at the instantaneous wavefunction $|\psi(t)\rangle$. %R3Q1prime
The time evolution of the ensemble-averaged density matrix is computed using the Krylov subspace method\,\cite{manmana2007strongly}.

\section*{Data Availability}
The numerical data that support the findings of this study are available from the corresponding authors upon reasonable request. 

\section*{Code Availability}
The relevant scripts to reproduce all figures are available from the corresponding authors upon reasonable request.

\bibliography{references,references_prev}

\begin{thebibliography}{72}
\bibitem [{pol()}]{polarizationSymmetry}%
  \BibitemOpen
  \href@noop {} {}\bibinfo {note} {Because the cluster preserves the $C_4$ symmetry, the direction of linear pump polarization does not introduce any directional bias in the analysis. In other words, switching the pump polarization from $x$ to $y$ leads to a straightforward interchange of the $\Phi_x$ and $\Phi_y$ outcomes.}\BibitemShut {Stop}%
\end{thebibliography}

\section*{Acknowledgements}
We acknowledge insightful discussions from Cheng-Chien Chen, Martin Claassen, and Matteo Mitrano. This work is supported by the Air Force Office of Scientific Research Young Investigator Program under grant FA9550-23-1-0153. The simulations presented in this paper were performed on the Frontera computation system at Texas Advanced Computing Center, operated under NSF Award No.OAC-1818253.

\section*{Author contributions} 
Y.W. conceived the project. Z.S. and W.-C.C. performed the calculations and data analysis. Z.S., C.X. and Y.W. wrote the manuscript.

\section*{Competing interests} The authors declare no competing interests.

\end{document}